\documentclass[prc,aps,showpacs,twocolumn]{revtex4}
\usepackage{amsmath}
\usepackage{amssymb}
\usepackage{graphicx}
\usepackage{color}
\makeatletter

\newcommand{\beq}{\begin{equation}}
\newcommand{\eeq}{\end{equation}}
\newcommand{\beqa}{\begin{eqnarray}}
\newcommand{\eeqa}{\end{eqnarray}}

\newcommand{\ket}[1]{\big| \,{#1}\, \big> }

\newcommand{\braket}[2]{\big< \,{#1}\, \big| \,{#2}\, \big> }

\newcommand{\expect}[1]{\big< \, {#1} \, \big>}

\newcommand{\matrixe}[3]{\big< \,{#1}\, \big| \,{#2}\, \big| \,{#3}\, \big> }

\newcommand{\op}[1]{%
    \fontdimen12\textfont3=2pt\fontdimen12\scriptfont3=1.4pt%
    \!\null\mathop{\vphantom{#1}\smash{#1}}\limits_{\!\sim}\null\!}

\newcommand{\vek}[1]{\!\vec{\,#1}}

\newcommand{\half}{{\textstyle \frac{1}{2}}}

\newcommand{\fm}{\mathrm{fm}}
\newcommand{\sheq}{\!=\!}
\newcommand{\UCOM}{\mathrm{UCOM}}

\makeatother
\begin{document}
\title{Long Range Tensor Correlations in\\
Charge and Parity Projected Fermionic Molecular Dynamics}

\author{Sonia Bacca\footnote{present address: TRIUMF, 4004 Wesbrook Mall, 
                              Vancouver B.C., V6T 2A3, Canada},   
Hans Feldmeier and  Thomas Neff }

\affiliation{Gesellschaft f\"{u}r Schwerionenforschung, Planckstr.~1,
64291 Darmstadt, Germany}


\begin{abstract}
Within the framework of Fermionic Molecular Dynamics a method is developed to 
better account for long range tensor correlations in nuclei when working with 
a single Slater determinant.
Single-particle states with mixed isospin and broken parity build up an intrinsic
Slater determinant which is then charge and parity projected.
By minimizing the energy of this many-body state with
respect to the parameters of the single-particle states
and projecting afterwards on angular momentum 
ground state energies are obtained that are systematically
lower than corresponding Hartree-Fock results.
The realistic Argonne V18 potential is used and short range
correlations are treated with the Unitary Correlation Operator Method.
Comparison with exact few-body calculations shows that in $^4$He  
about one fifth of the correlation energy due to long-range correlations are accounted for. 
These correlations which extend over the whole nucleus are visualized with 
the isospin and spin-isospin density of the intrinsic state.
The divergence of the spin-isospin density, the source for pion fields,
turns out to be of dipole nature.
\end{abstract}

\maketitle

\section{Introduction}

Despite the undeniable successes of mean field models a more microscopic view on the
structure of atomic nuclei reveals correlations among the nucleons of various kinds.
These are induced by realistic nucleon-nucleon interactions and cannot be 
represented by a single Slater determinant in a mean field picture.

Impressive progress has been made in the derivation of the nuclear interaction, 
one of the latest achievements being the nucleon-nucleon (NN) forces  and consistent 
many-nucleon forces developed with chiral perturbation theory
\cite{EnM03,epelbaum}. 
Furthermore, a number of other nuclear potentials exist, 
like the semi-phenomenological Argonne V18 potential (AV18) \cite{wiringa:1995}, 
which reproduce the experimental NN phase shifts with high precision.
Common to the realistic NN forces mentioned before are two general features:
(i) they are strongly repulsive at very small distances, preventing
nucleons to stay close together, which induces short-range repulsive radial correlations 
among them;
(ii) they contain a tensor force component, such that the nucleon pair feels a force
which depends on their spin orientation with respect to the relative distance.
This induces further short and long range tensor correlations among nucleons.

It is well known that a Slater determinant (antisymmetrized product state)
cannot represent such correlations. Models that use a Slater determinant basis need 
very large many-body Hilbert spaces to represent these correlations. 
Therefore in the no-core shell model (NSCM) \cite{NCSM} 
the unitary Lee-Suzuki transformation is incorporated  
to treat the short range part of the Hamiltonian that scatters
to very high lying oscillator shells and thus helps to improve convergence and to 
reduce the dimensions of the Hilbert space.

Another possible solution to overcome the problems caused by the short-range 
correlations is the Unitary  Correlation Operator Method (UCOM)  
\cite{feldmeier:1998, neff:2003, roth:2004}. 
A unitary correlator  $\op{C}=\op{C}_\Omega \op{C}_r$ is devised to imprint explicitly the 
short range tensor correlations, by means of $\op{C}_\Omega$, and radial correlations, 
by $\op{C}_r$, into the uncorrelated Slater determinant basis. 

However, tensor correlations are not only of short range like the radial repulsive 
correlations, but contain parts of long-range nature originating from the one pion 
exchange, exemplified e.g. by the large extension of the $D$-wave component of the 
deuteron wave function.
Therefore, even when using the UCOM approach long-range correlations need to be 
incorporated by configuration mixing in the Hilbert space.
In case of many-body methods based on an expansion of the states in terms of a complete 
set of  basis states one is in principle able to account for all kind 
of correlations, provided that convergence in the expansion is reached.
But when working with a low-momentum basis, like with just one or few
Slater determinant states, the incapability of describing long-range correlations
in nuclei still constitutes a big limitation.

In this work, we would like to address the general problem of how to account for
long-range tensor correlations within a restricted Slater determinant basis.
We propose a method, which follows  ideas presented  by 
Ikeda, Sugimoto and Toki published in a series of papers 
\cite{STI03,IST04,SIT04a,sugimoto,SIT07} to allow proton and neutron
states to mix at the single-particle level and then project the many-body state
on good charge and parity.
We introduce additional variational degrees of freedom in the
Fermionic Molecular Dynamics (FMD) framework which allow for
the possibility that the isospin of a nucleon can point in any direction
in isospin space and not just in the proton or neutron direction.

The paper is organized as follows.
In Sec.~\ref{Sec:Overview} the theoretical background is set:
an overview of the FMD and of its extension and the used effective interaction
are presented.
In Sec.~\ref{Sec:Results} results are shown
and  conclusions are finally drawn in Sec.~\ref{Sec:Conclusions}.

\section{Theoretical overview}
\label{Sec:Overview}
\subsection{Fermionic Molecular Dynamics}

In the following we will briefly introduce the Fermionic Molecular Dynamics (FMD) approach
and the generalization of the FMD wave functions to mixed charge states.
In FMD \cite{Fel89,FMD,neff:2004, neff:2005, FMD:neff, Varenna} the many-body Hilbert space is spanned by
non-orthogonal many-body basis states which are given by antisymmetrized products of 
single particle states, as
\begin{equation}
\ket{Q} =\op{\mathcal{A}}\Big( \ket{q_1} \otimes \dots  \otimes \ket{q_A} \Big) \ .
\label{SD}
\end{equation}
$\op{\mathcal{A}}$ is the antisymmetrization operator and $\ket{q_k}$ denote the 
single-particle states
which are linear combinations of Gaussian wave packets localized in phase-space, with 
variable spin orientation
$\ket{\chi}$ and the generalized isospinor $\ket{\xi}$, 
\begin{equation}
\braket{\vek{x}}{q}=\sum_i  c_i
\exp \left\{ -\frac{(\vek{x} - \vek{b}_{i})^2}{2 a_{i}}\right\}
\otimes \ket{\chi_i} \otimes \ket{\xi_i} \ .
\label{eq:paramFMD}
\end{equation}
Each Gaussian wave packet is parameterized in terms of a complex vector $\vek{b}_i$,
indicating the mean position and momentum and a complex width parameter $a_i$,
which can be different for each Gaussian, in contrast to the AMD approach \cite{AMD}
where the widths are real and common for all nucleons.

The spinor is  parameterized via two complex components for spin-up and spin-down
\begin{equation}
\ket{\chi_i}= \chi^{\uparrow}_{i}\; \ket{\uparrow} +\; \chi^{\downarrow}_{i}\; \ket{\downarrow}
\end{equation}
allowing for all orientations of the spin.
Analogously the generalized isospin part $\ket{\xi_i}$ describes a linear superposition 
of proton $\ket{p}$ and neutron $\ket{n}$
\begin{equation}
\ket{\xi_i}= \xi^{p}_{i}\; \ket{p} +\; \xi^{n}_{i}\; \ket{n}
\label{paramFMD2}
\end{equation}
so that a nucleon can adopt any ``direction'' in isospin space.
In former applications of FMD $\ket{\xi}$ was either a proton $\ket{p}$ or a neutron
state  $\ket{n}$.

This generalization introduces charge mixing in the single-particle and many-body space.
In a description, where the degrees of freedom are only nucleons and the charged
mesons do not appear explicitly, the charge carried by the nucleons is a sharp 
quantum number.
The many-body state $\ket{Q}$ however breaks the symmetry with respect to isospin 
rotations around the 3-axis. Therefore one has to project on the desired charge number
corresponding to the eigenvalue $M_T$ of the third component of the total isospin
\begin{equation}
 \op{T}^{(3)}=\sum_{k=1}^A\ \op{t}^{(3)}(k) =\frac{1}{2}\sum_{k=1}^A\ \op{\tau}^{(3)}(k)\ .
\end{equation}
This is achieved by the charge projection operator \cite{delta_ref}
\begin{equation}
\op{P}^{M_T}  =
\frac{1}{A}\sum_{n=1}^{A} \exp\Big\{i\frac{2\pi n}{A}(\,\op{T}^{(3)} - M_T)\Big\}
\ .
\label{eq:chargeprojector}
\end{equation}
Thus, the charge projected state,  can be written as a superposition of Slater determinants 
$\ket{Q^{(n)}}$ obtained by rotating the single determinant $\ket{Q}$ in isospin space about 
an angle $2\pi n/A$:
\begin{align}
\ket{Q;M_T} &= \op{P}^{M_T} \ket{Q}\nonumber\\
  &=\frac{1}{A}\sum_{n=1}^A e^{-i\frac{2\pi n}{A}M_T} \ket{Q^{(n)}} \ .
\end{align}

The following simple example shows that the projection of a two-body product state
with mixed charges results in correlated states with good isospin. From
the product states of two nucleons
\begin{align}
 \Big(\ket{p}\pm\ket{n}\Big)&\otimes\Big(\ket{p}+\ket{n}\Big) =\nonumber\\
  &\ket{p}\otimes\ket{p}\ \pm\  \ket{n}\otimes\ket{n}\nonumber\\
 &+\Big(\ket{p}\otimes\ket{n}\pm \ket{n}\otimes\ket{p}\Big)
\end{align}
one can project out all 4 components with isospin $T\sheq 1$ and $0$
including the $M_T=0$ two-body states
\begin{align}
 \ket{T\sheq 1, M_T\sheq 0}&=
            \frac{1}{\sqrt{2}}\Big(\ket{p}\otimes\ket{n}+ \ket{n}\otimes\ket{p}\Big)\\    
\ket{T\sheq 0, M_T\sheq 0}&=
            \frac{1}{\sqrt{2}}\Big(\ket{p}\otimes\ket{n} - \ket{n}\otimes\ket{p}\Big)     
\end{align}
which are correlated and cannot be written as
a product of two single-particle states.

As the lightest charged meson, the pion, is of pseudoscalar nature
with a negative intrinsic parity it may carry besides charge also parity from
one nucleon to the other.
Therefore we also allow for parity breaking in the FMD state, which simply means that
the parameters $\vek{b}_i$ are not the same for all nucleons, and restore it
by projection on good parity $\pi\sheq \pm 1$ with the projection operator
\begin{equation}
 \op{P}^{\pi} = \frac{1}{2}(\; \op{1}+\pi\ \op{\Pi}\;) \ ,
\end{equation}
where $ \op{\Pi}$ is the parity operator.

In order to see in how far Slater determinants, that break charge and parity,
can represent long range correlations induced by the exchange of pions
variational calculations are performed.  We minimize the energy
\begin{equation}
 \frac{\matrixe{Q;\pi,M_T}{\op{H}}{Q;\pi,M_T}}{\braket{Q;\pi,M_T}{Q;\pi,M_T}}=
  \frac{\matrixe{Q}{\op{H}\;\op{P}^{M_T}\op{P}^\pi}{Q}}{\matrixe{Q}{\op{P}^{M_T}\op{P}^\pi}{Q}}
\label{eq:energy}
\end{equation}
of the charge and parity projected FMD state
\begin{equation}
 \ket{Q;\pi,M_T}=\op{P}^{M_T}\op{P}^\pi\ket{Q}
\label{eq:QpiM_T} 
\end{equation}
with respect to all single-particle parameters contained in the single Slater determinant $\ket{Q}$.

As the intrinsic Hamiltonian $\op{H}$ commutes with $\op{\Pi}$ and $\op{T}_3$ one has
to project only the ket-state so that the number of terms in the energy \eqref{eq:energy} is reduced.

The correlated many-body state $\ket{Q;\pi,M_T}$ that results from the minimization
of the energy \eqref{eq:energy} in general breaks rotational and translational symmetry.
Therefore we project after the variation  on good angular momentum and
center of mass momentum zero \cite{FMD:neff,Varenna}:
\begin{equation}\label{eq:projstate}
 \ket{Q;J^\pi,MK,M_T}=\op{P}^J_{M\!K}\op{P}_{C\!M}\ket{Q;\pi,M_T}\ .
\end{equation}
Throughout this paper we consider only energies of $J^\pi=0$ states so that
$K=M=0$. All energies are calculated as expectation values of
the intrinsic Hamiltonian $\op{H}=\op{T}-\op{T}_{cm}+\op{V}$ where the
center of mass kinetic energy has been subtracted:
\begin{equation}
 E(0^+)=\frac{\matrixe{Q}{\op{H}}{Q;J^\pi\sheq 0^+,00,M_T}}{\braket{Q}{Q;J^\pi\sheq 0^+,00,M_T}}\ .
\label{eq:energyJproj} 
\end{equation}

It would be of course desirable to first restore all symmetries of the Hamiltonian and then 
do the variation with respect to the single-particle parameters contained in $\ket{Q}$.
However, angular momentum projection requires the superposition of a few hundred
rotated states, which is numerically very costly.
Therefore we do a variation after charge and parity projection and project on
angular momentum and CM-momentum zero after variation.

\subsection{Unitary Correlation Operator Method}

Realistic nucleon nucleon interations, like the chiral forces or the Argonne V18 
potential, that reproduce the phase shifts and the deuteron properties,
induce strong short range correlations. The repulsive core prevents two nucleons from
getting too close. In the $T\sheq 0$ channel the tensor force induces correlations 
by aligning the spins of two interacting nucleons along their distance vector.
Both correlations cannot be represented by Slater determinants.
Even though the Slater determinants of a shell model basis form a complete set,
diagonalization of a realistic Hamiltonian in the 4-body space shows that 
convergence cannot be reached in any tractable shell model space \cite{roth:2005}.
Thus many-body Slater determinants are an inadequate representation
for these short ranged correlations. 
In fact the no-core shell model employs the Lee-Suzuki transformation of the Hamiltonian 
which improves the convergence dramatically.

Here we will use the Unitary Correlation Operator Method (UCOM) to take care of the
short range correlations. 
The reason is that an effective interaction to be used in the FMD Hilbert space
has to be represented in an operator form, rather than in terms of matrix elements
as is the case for a Lee-Suzuki transformation or a G-matrix. UCOM provides both
representations so that we can compare with no-core shell model results.

The concept of UCOM \cite{feldmeier:1998, neff:2003, roth:2004} consists in treating
explicitly the strong short range correlations induced by the hard core and the short 
range part of the tensor force 
by a state independent unitary transformation, while the long range correlations
have to be represented by the many-body states spanning the Hilbert space.  
When the correlation operator is applied to an initial Hamiltonian, a phase-shift
equivalent correlated interaction is obtained.
The correlated Hamiltonian is defined via a similarity transformation
\begin{align}
\op{\widehat{H}}=&\ \op{C}^{-1}_{r}\op{C}^{-1}_{\Omega}\op{H}_{\rm initial}\,
\op{C}_{\Omega}\,\op{C}_{r}
\nonumber\\
=&\ \op{T}-\op{T}_{cm}+\op{V}_\UCOM+3{\rm -body}+\dots
\label{eq:correlated_H}
\end{align}
where $\op{C}_{r}$ and $\op{C}_{\Omega}$  are  the unitary central
and tensor correlation operators, respectively.
$\op{C}_{r}$ shifts pairs of nucleons radially away from each other whenever
their distance is so small that they would be inside the repulsive core.
$\op{C}_{\Omega}$ aligns nucleon pairs with $T\sheq 0$ and $S\sheq 1$ along 
the direction of their total spin so that the typical tensor correlations
known from the deuteron are imprinted into the many-body state.
Different from the radial correlations induced by the short ranged
repulsion, the tensor interaction mediated by the exchange of pions,
the lightest of all mesons, induces long range correlations. 
In order to keep the effect of the induced many-body interactions small
one restricts the range of the action of the tensor 
correlation operator $\op{C}_{\Omega}$ and represents the
long range part of the tensor correlations with the 
many-body state (see Ref.~\cite{roth:2005} for details).

The two-body part of the correlated Hamiltonian \eqref{eq:correlated_H}
is used as an effective intrinsic Hamiltonian 
\begin{equation}
\op{H}=\op{T} - \op{T}_{cm}+ \op{V}_\UCOM
\label{eq:Heff} 
\end{equation}
that is applicable in low momentum Hilbert spaces. The effective
interaction $V_\UCOM$ is phase shift equivalent to the initial
realistic interaction. Moreover different realistic potentials, like Bonn A,
Nijmegen or Argonne V18, lead to practically the same $V_\UCOM$. 
In FMD the effective potential needs to be given in operator
representation and not as momentum space matrix elements for 
the different partial waves. $V_{\UCOM}$ can easily be obtained in 
operator form if the initial interaction is given in terms of operators.
Therefore we use in this
publication the Argonne V18 potential (AV18) as initial potential
because it is already in operator form.

For a better understanding of the results let us first consider in somewhat more
detail how the tensor correlator $\op{C}_\Omega$ acts. Any operator that 
depends only on the distance $r$ between two nucleons is invariant under
actions of the tensor correlator
\begin{equation}
\op{C}_\Omega^{-1}\;V(\op{\,r})\;\op{C}_\Omega=
            e^{+i\op{g}_{\Omega}}\;V(\op{\,r})\;e^{-i\op{g}_{\Omega}}=V(\op{\,r})\ ,
\end{equation}
because the tensorial generator
\begin{equation}
\op{g}_\Omega=\vartheta(\,\op{r})\frac{3}{2}\Big(
   (\,\op{\vek{p}}_\Omega\!\cdot\!\op{\vek{\sigma}}_1)
                            (\,\op{\vek{r}}\!\cdot\!\op{\vek{\sigma}}_2)+
   (\,\op{\vek{r}}\!\cdot\!\op{\vek{\sigma}}_1)
                            (\,\op{\vek{p}}_\Omega\!\cdot\!\op{\vek{\sigma}}_2)
    \Big)
\end{equation}
commutes with $\op{r}=|\op{\vek{r}}|$. 
The reason is that the so called orbital momentum $\op{\vek{p}}_\Omega$, 
defined as the component of the relative momentum perpendicular to $\op{\vek{r}}$, 
\begin{equation}
\op{\vek{p}}_\Omega=\op{\vek{p}}\ -\ \op{p}_r\;\frac{\op{\vek{r}}}{\op{r}}  \ ,
 \end{equation}
commutes with $\op{r}$.
The correlation function $\vartheta(r)$ defines the strength of the transformation
as a function of the distance between an $S\sheq 1$ nucleon pair.

In the following we shall use different $\vartheta(r)$ (displayed in Fig. \ref{fig:1})
with varying range characterized by the range parameter
$I_\vartheta=\int dr\ r^2 \vartheta(r)$. 

When applying the tensor correlator to the tensor interaction 
\begin{equation}\label{eq:tensorpotential}
V_T(\op{r})\op{S}_{12}=V_T(\op{r})\;
\Big(\;\frac{3}{\op{r}^2}(\,\op{\vek{r}}\!\cdot\!\op{\vek{\sigma}}_1)
(\,\op{\vek{r}}\!\cdot\!\op{\vek{\sigma}}_2)
-(\op{\vek{\sigma}}_1\!\cdot\!\op{\vek{\sigma}}_2)\Big)
\end{equation}
it is unitarily ``rotated'' 
into a reduced tensor $\tilde{{V}}_{T}$, a central $\tilde{{V}}_C$, and a spin-orbit
 $\tilde{{V}}_{LS}$ component, plus some other terms which are negligibly small.
\begin{eqnarray}\label{eq:correlation_on_tensor}
\op{C}_\Omega^{-1}\;V_T(\op{r})\op{S}_{12}\; \op{C}_\Omega  
&=&\op{\tilde{V}}_T+\op{\tilde{V}}_C+\op{\tilde{V}}_{LS}+\cdots\\
\nonumber &=& e^{-3\vartheta(\,\op{r})}\ \ V_T(\op{r})\ \ \op{S}_{12}\\
\nonumber &+&2(1-e^{-3\vartheta(\,\op{r})})\ V_T(\op{r})\ (3+\op{\vek{\sigma}}_1 \!\cdot\! 
\op{\vek{\sigma}}_2)\\
\nonumber &+&6(1-e^{-3\vartheta(\,\op{r})})\ V_T(\op{r})\ \op{\vek{l}} \!\cdot\! 
\op{\!{\vek{s}}} \\
\nonumber &+& \cdots 
\end{eqnarray}
In Eq.~\eqref{eq:correlation_on_tensor} one readily sees that the induced
central force  $\tilde{{V}}_C$ increases if the correlation function
$\vartheta(r)$ increases.
Thus whith larger range of the tensor correlator more strength goes
to the central and spin-orbit interaction. 
One should keep in mind that
the correlation function $\vartheta(r)$ is different for 
isospin $T=0$ and $T=1$, see Ref.~\cite{neff:2003}
and that there are also contributions from the 
correlated kinetic energy $\op{C}_\Omega^{-1}\;\op{T}\; \op{C}_\Omega$,
the spin orbit force 
\mbox{$\op{C}_\Omega^{-1}\;\op{\vek{l}}\!\cdot\!\op{\!{\vek{s}}}\;\op{C}_\Omega$}
and possibly other terms in the interaction.  

Eq.~\eqref{eq:correlation_on_tensor} also gives finally an 
explanation why in most textbooks after the discussion of the deuteron
with its quadrupole moment and the mandatory tensor force
the tensor interaction dissappears in the following
chapters on mean-field theories like Hartree-Fock or 
simple shell-model pictures.
The reason is that those calculations can be performed without the
tensor force using only central and spin-orbit interactions
that are fitted to energies and other properties of the many-body system.
The success of these models does not imply that in finite nuclei
or nuclear matter the tensor force is not important. 
By fitting the parameters of the interaction one has effectively
moved the correlation energy of the tensor into central and spin-orbit
interactions. However, the resulting effective interactions are
not phase-shift equivalent any longer, $V_\UCOM$ is. Furthermore,
the uncorrelated many-body Slater determinants do not possess
the tensor correlations anymore that are existent in reality and
can be observed for example in the high momentum part of the momentum distribution
of the nucleons \cite{neff:2003}.

\section{Results}
\label{Sec:Results}
In the following subsections we study in how far long range correlations
originating from the tensor interaction  
can be repesented by a charge and parity projected Slater determinant.
For that the realistic AV18 interaction is used and the results are compared to
exact solutions for the 4-body system $^4$He.

In subsection \ref{Sec:sugimoto} we use a phenomenological interaction 
that is not based on a realistic nucleon nucleon force to illustrate that
fitting interaction parameters to a specific nucleus may be misleading
when drawing conclusions about the strength of correlations.

\subsection{Correlation energies}
\label{sec:correlationenergy}

In this section we investigate how the induced correlations are accounted for by breaking 
charge and parity in each single-particle state as a function of the strength of 
the tensor component of the nucleon-nucleon interaction.
In order to get a fair judgement we use the realistic effective
interaction $V_\UCOM$ which is derived from the AV18 potential.
The strength of its tensor component depends on the range $I_\vartheta$ 
of the tensor correlator. This range corresponds to the cutoff in momentum space in the 
$V_{{\rm low}\,k}$-approach \cite{Vlowk, neff:2003}.
For all values of $I_\vartheta$ the effective interaction $V_\UCOM$
is phase shift equivalent to the original realistic interaction.
It should be noted that for different $I_\vartheta$
all parts of the $V_\UCOM$ potential change accordingly 
under the similarity transformation \eqref{eq:correlated_H}
and not just the tensor part 
(see also Eq.~\eqref{eq:correlation_on_tensor}).

We calculate the ground state energy of $^4$He in three different ways
using the same effective Hamiltonian \eqref{eq:Heff} for different $I_\vartheta$. 

First we minimize the energy $\matrixe{Q}{\op{H}}{Q}/\braket{Q}{Q}$ 
with respect to all FMD parameters contained in $Q$ without any projection.
The FMD single-particle states $\ket{q_k}$ contain two Gaussians, 
see Eq. \eqref{eq:paramFMD}.
The results are labelled HF as this variation corresponds to a Hartree-Fock minimization.
Second we minimize the energy \eqref{eq:energy} of the charge and parity projected
FMD Slater determinant \eqref{eq:QpiM_T}, labelled CPP. 

After that an angular momentum and CM projection is applied to the
HF and CPP state, see Eq.~\eqref{eq:projstate}. The HF state has already 
$J^\pi\sheq 0^+$ and $M_T=0$ but the CPP state has no good angular momentum. 
The resulting energies \eqref{eq:energyJproj} are compared to the exact results
of no-core shell model and hyperspherical harmonics calculations.

\begin{figure}[t]
\begin{center}
\includegraphics*[scale=0.6]{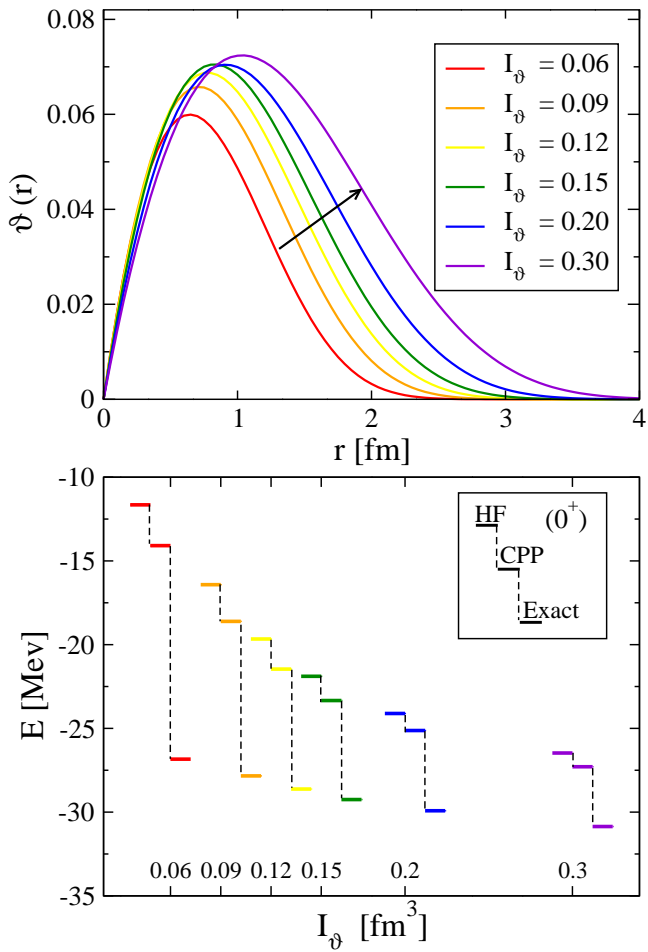}\\[1ex]
\end{center}
\caption{\label{fig:1}(Color online)
Upper panel: tensor correlation functions $\vartheta(r)$ with increasing 
range $I_\vartheta$ (arrow).
Lower panel: ground state energies of $^4$He calculated with the HF and CPP states
compared to exact results. The correlation functions shown in the upper panel
are used. }
\end{figure}

In order to study the effects of different tensor strengths in the Hamiltonian
we vary the range of the tensor correlator in the $S\sheq 1$, $T\sheq0$ channel,
which according to Eq.~\eqref{eq:correlation_on_tensor} moves 
strength from the tensor to the central part. This has the advantage
that we are always using phase-shift equivalent potentials. In the upper
part of Fig.~\ref{fig:1} we display the correlation functions employed.
They differ in range which we quantify by the range parameter 
$I_\vartheta$.

In the lower part of Fig.~\ref{fig:1} we present the binding energies of $^4$He
obtained with the $V_{\UCOM}$ potential for different $I_{\vartheta}$
in the three variational Hilbert spaces as explained before.
The range $I_{\vartheta}=0.09\ \fm^3$ corresponds to the standard
choice where $V_\UCOM$ reproduces within 0.5 MeV both the $^4$He and $^3$He binding
energies. This interaction has been used for many applications, 
ranging from few- to many-body problems 
\cite{roth:2005, roth:2006, paar:2006, carlo:2006, sonia_ucom}.
The range $I_{\vartheta}=0.09\ \fm^3$ is optimal in the sense that the 
net contribution from 3-body forces is minimized \cite{roth:2005}.
 
The exact results in Fig.~\ref{fig:1} show that ranges smaller than
$I_\vartheta$=0.09~fm$^3$ lead to underbinding with respect to
the experimental energy of -28.3 MeV 
(like for most realistic potentials). 
On the other hand tensor correlators with ranges larger than $I_\vartheta$=0.09~fm$^3$
induce 3- and 4-body interactions
in the correlated Hamiltonian $\widehat{H}$, Eq.~\eqref{eq:correlated_H},
which are on average repulsive but not included here.
A similar kind of overbinding was obtained with $V_{{\rm low}\,k}$ for
cutoffs $\sim 1.6~{\rm fm}^{-1}$\cite{Vlowk}.

The first result to be observed in Fig.~\ref{fig:1} is the decreasing
difference between the HF energy and the exact result when the
range $I_{\vartheta}$ of the tensor correlator is enlarged. 
This can be understood if one keeps in mind that the HF state of $^4$He is a 
pure $(0s)^4$ configuration so that the tensor interaction cannot contribute.
All binding comes from the central part of the interaction.
According to Eq.~\eqref{eq:correlation_on_tensor} increasing the
correlation strength implies that the induced central part of the 
correlated tensor interaction \eqref{eq:correlation_on_tensor}
in the $S\sheq 1$ $T\sheq0$ channel 
increases and binding energy is gained, while the long range part of the
tensor force, not seen by the HF state, is reduced. 
The difference between the HF and the exact energy is the correlation
energy which is due to the interaction induced correlations present in the exact
many-body state but absent in the HF Slater determinant. 

The exact results obtained in an ``unrestricted'' Hilbert space show
much less variation as function of $I_\vartheta$ because there the long range
correlations are represented in the many-body state even for small
$I_\vartheta$. If we would include the induced 3- and 4-body potentials
the exact energy would not depend on $I_\vartheta$ because the transformation 
would then be unitary even in the many-body space.

An important finding is that, although the variational manifold $\{Q\}$
includes the new isospin mixing degrees of freedom, the Hartree-Fock (HF)
minimum does not make use of them. All $\xi^p_i$ and $\xi^n_i$ are either
0 or 1, so that the Hartree-Fock state is already an eigenstate of charge. 
It turns out to be also an eigenstate of angular momentum and parity $(J^\pi=0^+)$.

This changes when the variation is performed after charge and parity projection.
Now the $\xi^p_i$ and $\xi^n_i$ parameter assume values different from 0 or 1.
As Fig.~\ref{fig:1} shows, 
the charge and parity projected state (CPP) can represent part of the 
long range correlations leading to a correlation energy of about 20\% the full
correlation energy. 

In Table \ref{tab:1} the expectation values of the total Hamiltonian, the kinetic energy, 
the interaction energy and the tensor part of the correlated interaction 
$V_\UCOM$ are listed.
%
\begin{table}
\caption{\label{tab:1}
Expectation values of different terms of the Hamiltonian for the 
$^4$He $J^\pi=0^+$ ground state.
The variation of single-particle states is performed for 
Hartree-Fock (HF), Parity Projected (PP) and Charge and Parity Projected (CPP)
intrinsic Slater determinants.  
$V_{\UCOM}$ with tensor correlator range $I_{\vartheta}=0.09$ fm$^3$ is used.
Numerical values are in MeV.}
\begin{ruledtabular}
\begin{tabular}{crrr}
{}&{HF[0$^+$]}&{PP [0$^+$]} &{CPP [0$^+$]}\\
\hline
$\expect{\op{H}}$ & -16.42 & -17.51 &  -18.61\\
$\expect{\op{T}}$ &  50.25	&  51.56 &   57.93\\
$\expect{\op{V}}$ & -66.68 & -69.07 &	-76.54\\
$\expect{\op{V}_T}$ & 0.00 &  -1.35 &   -4.59\\
\end{tabular}
\end{ruledtabular}
\end{table}
The HF state has zero tensor energy and the smallest kinetic energy, while the
CPP state contains correlations that yield about $-2.2$~MeV more binding. 
These consist of $7.7$~MeV kinetic energy counterbalanced by $-9.9$~MeV additional
potential energy. Out of that the tensor contributes $-4.6$~MeV. The rest
originates from other parts. 
The rather large increase in kinetic energy is partly due to the fact that
in the CPP minimum the two Gaussians in the single-particle states
are displaced from each other, thus break parity and include higher
angular momenta. 
Therefore the spatial and spin parts are different in the CPP minimum and 
the HF minimum.

It is interesting to note that a variation
after parity projection (PP) already takes into account part of the correlations
as can be seen from the center column of Table~\ref{tab:1}.
One should also keep in mind that
with respect to an uncorrelated Slater determinant a by far larger amount of 
correlation energy resides in the short range
tensor correlations which are treated here by the 
Unitary Correlation Operator Method (UCOM). These short range correlations 
are so strong that they even cannot be properly accounted for in a 
very large scale shell model basis, not to mention a single Slater determinant.

The effects are not as pronounced if one restricts the single-particle
states to one Gaussian. Adding the second Gaussian is essential in describing tensor 
correlations, as it allows to add higher angular
momenta components to the single-particle wave functions.

\begin{figure}[t]
\begin{center}
\rotatebox{0}{\includegraphics*[scale=0.6]{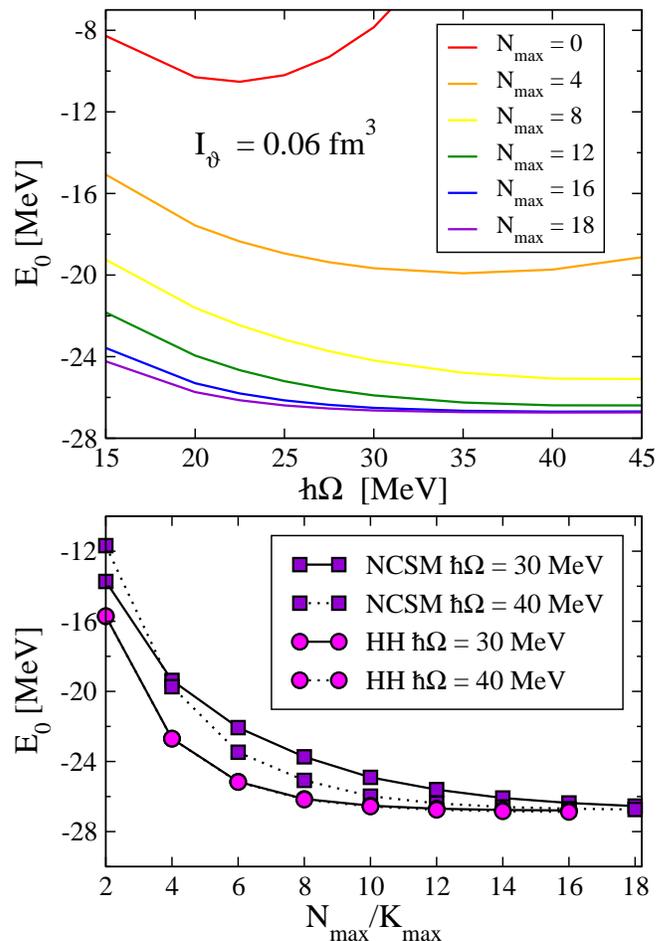}}
\caption{\label{fig:convergence}(Color online)
Convergence patterns for the NCSM calculations with the UCOM potential for 
$I_{\vartheta}=0.06$ fm$^3$: the ground state energy of $^{4}$He is plotted as a function 
of the oscillator parameter $\hbar\Omega$ for different model space sizes (upper panel). 
Comparison of the NCSM and HH convergence as a function of $N_{\rm max}$ and 
$K_{\rm max}$, 
respectively, for fixed  $\hbar\Omega$ (lower panel). }
\end{center}
\end{figure}
\begin{table*}
\caption{\label{table3}
Exact ground state energies (in MeV) of $^4$He for different ranges of the 
tensor correlator in UCOM, calculated with the NCSM and HH approaches.}
\begin{ruledtabular}
\begin{tabular}{ccccccc}
{$I_{\vartheta}$ [fm$^3$]}&0.06&0.09&0.12&0.15&0.20&0.30\\
\hline
NCSM&-26.80(3)&-27.80(3)&-28.62(9)&-29.26(13)&-29.86(10)&-30.70(6)~\\
HH&-26.84(6)&-27.84(7)&-28.62(7)&-29.25(8)&-29.92(10)&-30.86(10)\\
\end{tabular}
\end{ruledtabular}
\end{table*}

One should pay attention to the fact that additional correlations in the 
many-body state that are induced by the two-body interaction give an attractive
contribution from the potential but at the same time the kinetic energy rises. 
For example tensor correlations imply admixtures of higher angular momenta 
and thus more kinetic energy.
Hence the correlation energy is the result of a subtle interplay between enhanced 
kinetic energy and increased attractive potential energy. 

The ground state energies labelled ``Exact'' are calculated within the
no-core shell model (NCSM) using the translationally invariant harmonic oscillator
formulation of Petr Navr{\'a}til \cite{NCSM2}
and the hyperspherical harmonics (HH) approach developed by Nir Barnea
\cite{barnea:1997-98, barnea:1999}.
The latter method has been applied to the calculation of exact electroweak
reactions on light nuclei mainly with local interactions
\cite{GBB06,GBB06a,BMB02,BBL04,BAB04,Gaz08,BAB07}.
Recently it has been extended to use interactions represented in harmonic
oscillator (HO) states (see  e.g. Refs. \cite{sonia_ucom,barnea_JISP}).
The HH approach is equivalent to the NCSM, 
but makes use of HH functions instead of HO eigenstates.
The label ``Exact'' is meant in the sense that one systematically enlarges
the Hilbert space and controls the convergence to approach the
final many-body state which should contain all kind of correlations
induced by the interaction.

Convergence patterns for the NCSM calculations for a specific value of 
$I_{\vartheta}=0.06$~fm$^3$ are shown in Fig.~\ref{fig:convergence}. 
For a given size of the model
space, characterized by the maximum oscillator quantum number $N_{\rm max}$, the ground 
state energy is plotted as a function of the oscillator parameter $\hbar\Omega$. 
Convergence, resulting in a flat energy curve over a significant range of oscillator
parameters, can be obtained already with $N_{\rm max}=18$. The energy gained compared with 
the results obtained with more moderate model spaces can be attributed to the residual
long range correlations not described via the UCOM.

In Fig.~\ref{fig:convergence} we also compare the convergence of the HH and 
NCSM calculations for two fixed values of  $\hbar \Omega$ as a function  
of the $K_{\rm max}$ parameter, 
which is for the HH the analog to $N_{\rm max}$ for the NCSM. As already pointed out 
in Refs.~\cite{sonia_ucom} and \cite{benchmark} the convergence of HH is superior to 
the one of the NCSM,  since  no $\hbar \Omega$ dependence is observed even for small 
Hilbert spaces. Therefore we employ an exponential fit of the NCSM energies
as a function of $N_{\rm max}$ (for fixed $\hbar \Omega$) 
to extrapolate to infinite dimensions.
The further energy gain is of the order of $50-100$ keV.
The  energies obtained for different values of the tensor correlator volume 
$I_{\vartheta}$ used in the UCOM are shown in Table~\ref{table3}. 
The NCSM and HH results nicely agree with each other within the error bars.

The small difference of about 0.5~MeV we find in our exact calculations for 
$I_{\vartheta}=0.09$ fm$^3$ with
respect to the values previously published in Refs. \cite{roth:2005}
with the NSCM and \cite{sonia_ucom} with HH for the same $I_{\vartheta}$ is related to some
minor differences of the potential used. They originate from the fact that
in this paper  we adopt a  $V_{\UCOM}$ potential in operator representation which is
suited for the FMD code. This implies a truncation of the Baker-Campbell-Hausdorff expansion
(for details see Ref.~\cite{roth:2005}), not needed if one correlates directly the 
two-body states.

Here, we would like to stress again that the exact approaches make use of
thousands of basis states to reach convergence in energy, whereas with our improved 
FMD wave function we vary the parameters of a single Slater determinant 
projected on charge and parity.

\subsection{Pseudo-scalar iso-vector correlations}

A single Slater determinant can usually not represent correlations except
those induced by the Pauli principle.
However, if the Slater determinant represents an intrinsic state
the situation is different. 
The physical state that has the same symmetries as the Hamiltonian
is obtained by means of projection 
on angular momentum, parity and, as in our case, on charge. 
It is in general not a single Slater determinant but a superposition of many, 
see Eqs.~(\ref{eq:QpiM_T}, \ref{eq:projstate}), with the restriction that
all these Slater determinants are generated from a single one
by means of rotations, parity inversion or rotation in isospin space.

A well known example for long range correlations in nuclei are  
intrinsically deformed nuclei where out of one deformed intrinsic state
one projects a whole rotational band.

The intrinsic state for $^4$He obtained in this paper by variation after
parity and charge projection shows a pronounced long range correlation
in the spin isospin degrees of freedom. 
To elaborate on that let us consider the exchange of the
pseudo-scalar iso-vector pion.  

In all realistic nucleon-nucleon interactions the one-pion exchange is responsible
for the longe range tail of the potential. This part is not affected by the
unitary correlator $\op{C}_\Omega$ which is of short range. The induced long range
correlations should therefore be represented by the many-body state. 

The vertex describing the interaction of a nucleon field $N({\bf x})$
with a pion field $\Phi_{\pi}^{(i)}({\bf x})$ has in pseudo-vector 
coupling the following form
\begin{equation}
\mathcal{L}_{N\pi}({\bf x})=-\frac{g_{\pi}}{2M}\sum_{i=1}^3
\bar{N}({\bf x}) \gamma^5 \gamma_{\mu} \tau^{(i)} N({\bf x})
\partial^{\mu} \Phi_{\pi}^{(i)}({\bf x}) \ ,
\end{equation}
where $\tau^{(i)}, i=1,2,3$ denote the Pauli matrices in isospin space.
\begin{equation}
\tau^{(1)}\!=\!\left(\begin{array}{cc}0& 1 \\1&0\end{array}\right),\ 
\tau^{(2)}\!=\!\left(\begin{array}{cc}0&-i \\i&0\end{array}\right),\ 
\tau^{(3)}\!=\!\left(\begin{array}{cc}1& 0 \\1&-1\end{array}\right)
\end{equation}
In the stationary case the three components of the pion field 
$\Phi_{\pi}^{(i)}(\vek{x}\/)$ satisfy the 
time-independent Klein-Gordon equation 
\begin{equation}\label{eq:KG}
( -\nabla^2 + \ m^2_{\pi}  )\  \Phi_{\pi}^{(i)}(\vek{x}) =
\frac{g_{\pi}}{M}\  \vek{\nabla}\! \cdot\! \vek{S}^{(i)}(\vek{x})\ ,
\end{equation}
where the source term is the divergence of the 
nuclear isospin current density
\begin{equation} 
\vek{S}^{(i)}(\vek{x})=\frac{1}{2}\bar{N}(\vek{x})\, \gamma^5 \,
         \vek{\gamma}\, \tau^{(i)}\, N(\vek{x})\ .
\end{equation}
%
After a non-relativistic reduction to leading order 
${\vek{S}}^{(i)}(\vek{x})$  
becomes the one-body spin-isospin density of the nuclear 
many-body system
\begin{equation}
\op{\vek{S}}^{(i)}(\vek{x})
  =\frac{1}{2}\sum_{k=1}^A {\delta^3(\vek{x}-\op{\vek{r}}(k)) 
    \  \op{\vek{\sigma}}(k)\ \op{\tau}^{(i)}(k)} \ .
\end{equation}
The relation with the physical pion fields are
\begin{equation}
\begin{array}{ccc}
\pi^+(\vek{x})&=&\frac{1}{\sqrt{2}}\big(\Phi_{\pi}^{(1)}(\vek{x})+i\Phi_{\pi}^{(2)}(\vek{x})\big)\\
\pi^-(\vek{x})&=&\frac{1}{\sqrt{2}}\big(\Phi_{\pi}^{(1)}(\vek{x})-i\Phi_{\pi}^{(2)}(\vek{x})\big)\\
\pi^0(\vek{x}) &=& \Phi_{\pi}^{(3)}(\vek{x})
\end{array}\ .
\end{equation}
The (3)-component $\vek{S}^{(3)}(\vek{x})$ is the difference between the proton
and neutron spin density at position $\vek{x}$
and 
\mbox{$\vek{\nabla}\!\cdot\!\vek{S}^{(3)}(\vek{x})$}
is the pseudo-scalar iso-vector source density for the $\pi^0$ field, 
while
$\vek{\nabla}\!\cdot\!\vek{S}^{(1)}(\vek{x})$ and $\vek{\nabla}\!\cdot\!\vek{S}^{(2)}(\vek{x})$
are the sources for $\pi^\pm$ fields.

\begin{figure*}[htbp]
	\centering
\includegraphics[scale=0.678]{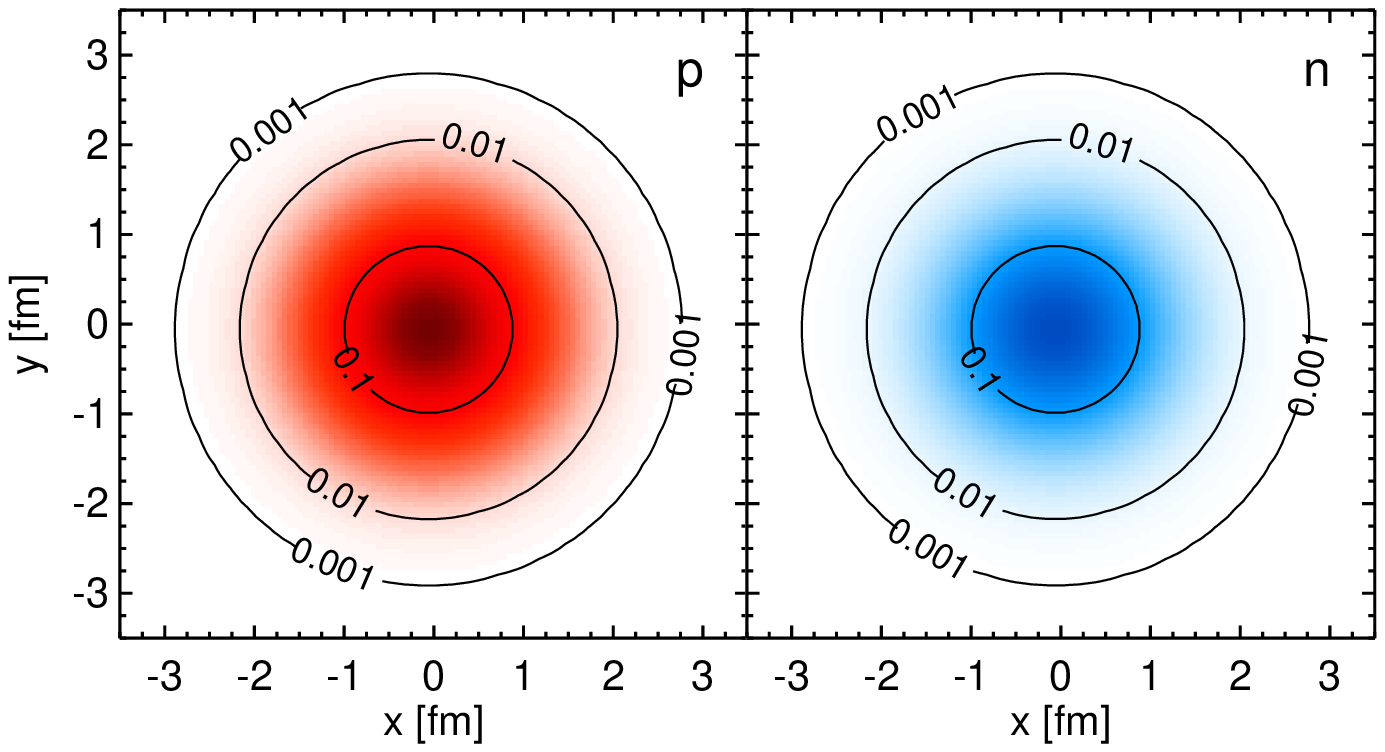}\\
\includegraphics[scale=0.75]{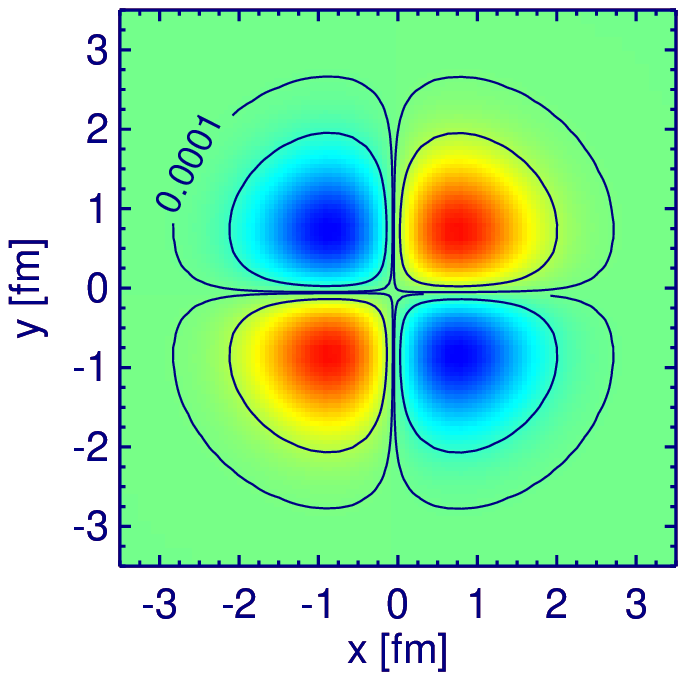}
\includegraphics[scale=0.75]{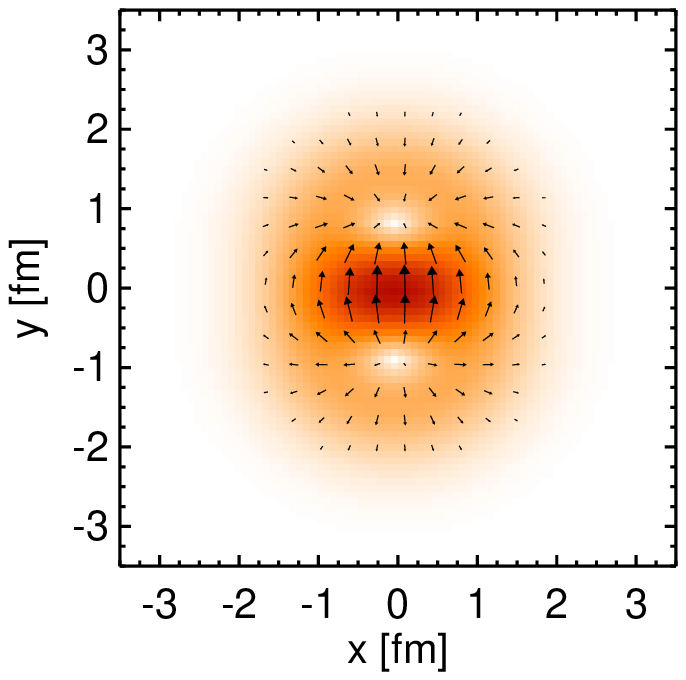}
\includegraphics[scale=0.75]{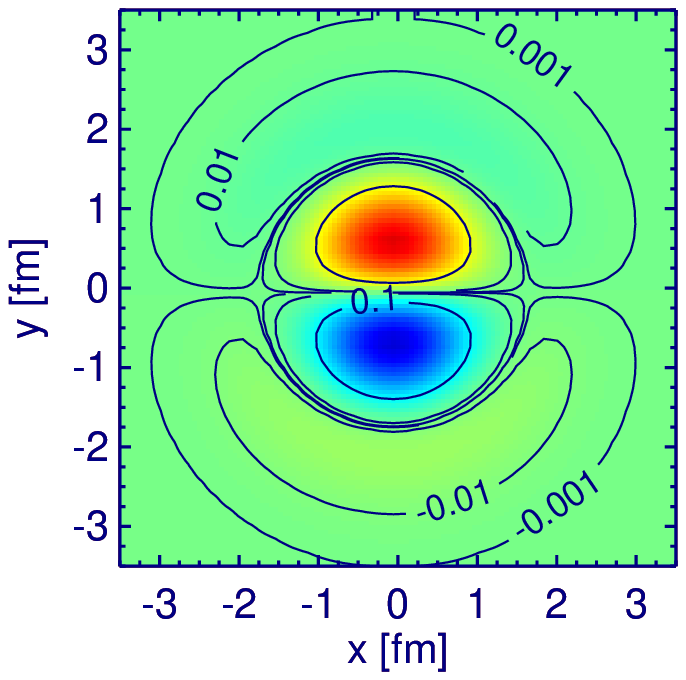}
	\caption{\label{fig:spin-isospin}(Color online)
	Densities calculated with parity projected intrinsic state $\ket{Q;+}$
	obtained for $I_\vartheta=0.09$ fm$^3$.\\
	Upper panel: proton density $\rho_p(\vek{x})$ 
	             and neutron density $\rho_n(\vek{x})$ (in fm$^{-3}$);
	lower panel: isospin density $\rho_\tau^{(1)}(\vek{x})$
	                  (in fm$^{-3}$); 
	             spin-isospin density
	             $\vek{S}^{(1)}(\vek{x})$;
	             divergence of spin-isospin density 
	             $\vek{\nabla} \cdot \vek{S}^{(1)}(\vek{x})$ (in fm$^{-4}$). 
	Contour lines in decades.
             }
\end{figure*}

In Fig.~\ref{fig:spin-isospin} various intrinsic densities are displayed
as a function of $x$ and $y$ at the plane $z=0$. All densities
are calculated with the parity projected intrinsic state
\begin{equation}
\ket{Q;+}:=\op{P}^{\pi=+1} \ket{Q}
\end{equation}
that was obtained by minimizing the 
energy after parity and charge projection (CPP), cf. Eq.~\eqref{eq:energy}.

Before we come to the spin-isospin density that is related to the 
pion fields we show in Fig.~\ref{fig:spin-isospin} the proton and
neutron densities for point-like nucleons. Both are equal up to very
small deviations caused by the Coulomb interaction. This means that
the isospin density 
$\rho^{(3)}_\tau (\vek{x})=\rho_p(\vek{x})-\rho_n(\vek{x})$
in the isospin $(3)$-direction is zero.
However, the isospin density 
\begin{align}
\rho^{(1)}_\tau (\vek{x})=
  \displaystyle{   
  \frac{\matrixe{Q;+}{\sum_{k=1}^A\delta^3(\vek{x}-\vek{\op{r}}(k))
      \op{\tau}^{(1)}(k)}{Q;+}}
                 {\braket{Q;+}{Q;+}}
                 }
\end{align}
in the $(1)$-direction orthogonal to the proton/neutron $(3)$-direction 
assumes a non-zero value. As seen in Fig.~\ref{fig:spin-isospin}
this density is of quadrupole type. 
The intrinsic state $\ket{Q;+}$ is apparently not an eigenstate of 
 $\;\op{T}^{(3)}$ and
its new degrees of freedom, $\xi^p,\xi^n$, that allow to mix protons
and neutrons, are responsible for this density.
The charge and parity projected state 
$\ket{Q;\pi=+1,M_T=0}$ has (like the Hartree-Fock state) again
a vanishing isospin-(1) density.

The lower panels in Fig.~\ref{fig:spin-isospin} display the spin-isospin density
\begin{equation}
\vek{S}^{(1)}(\vek{x})=
     \frac{\matrixe{Q;+}{\op{\vek{S}}^{(1)}(\vek{x})}{Q;+}}{\braket{Q;+}{Q;+}}
\end{equation}
and its divergence 
$\vek{\nabla} \cdot \vek{S}^{(1)}(\vek{x})$
which is the source density for the $\Phi^{(1)}_\pi(\vek{x})$ pion field.
$\vek{S}^{(1)}(\vek{x})$ represents a pseudo-vector 
iso-scalar field ((1)-component) with a pronounced dipole shape. 
The divergence is the according pseudo-scalar iso-vector source density. 
One should note that the structure of the intrinsic state extends over the 
whole nucleus and is hence of long range.

All other spin-isospin densities are two orders of magnitude smaller 
which means zero within numerical uncertainty and hence not displayed.
One should keep in mind that the intrinsic state $\ket{Q;+}$ can be rotated in isospin
space around the (3)-axis resulting only in an overall phase of the ground state, because
the intrinsic state is projected on good charge number by summing up rotations around
the (3)-axis, cf. Eq.~\eqref{eq:chargeprojector}.
Likewise one can rotate the intrinsic state in coordinate space without affecting
the angular momentum projected $0^+$ ground state, so that  
the dipole in $y$-direction could also point in any other direction.

Let us try to explain the physical meaning of the non-zero intrinsic 
pseudo-scalar isovector source density 
\mbox{$\vek{\nabla}\!\cdot\!\vek{S}^{(1)}(\vek{x})$}
with help of an analogy to the Coulomb interaction. 
Consider a positronium,
negativly charged electron plus positivly charged positron, in their
atomic ground state. This state has angular momentum zero and the 
probability to find a positron at some position equals that of the electron.
Hence the mean value or expectation value of the charge density $\rho_e$ is zero
and consequently there is no Coulomb field $\Phi_e$ which of course must not be
interpreted that there is no Coulomb attraction.
But if we take a ``snap shot'' we find the positron and electron on
opposite sides (perfect correlation) forming a dipole with non-zero charge density and 
non-zero Coulomb field by which they attract each other. After projecting
this dipole on a $0^+$ state we get the true positronium ground state.

Looking at Eq.~\eqref{eq:KG} and replacing
\begin{equation}\nonumber
\begin{array}{rcl}
\Phi^{(i)}_\pi         &\rightarrow&\Phi_e \\
\displaystyle{\frac{g_\pi}{M}}\vek{\nabla}\!\cdot\!\vek{S}^{(i)}&\rightarrow& 
4 \pi \rho_e\\ 
m_\pi &\rightarrow& 0
\end{array}
\end{equation} 
one recovers the well known equation for a Coulomb field created by a charge
density. 
The analogy is obvious, the intrinsic state is the ``snap shot'' where
one sees the dipole like source density as a one-body mean field 
(Fig.~\ref{fig:spin-isospin}).
If one wants to see that correlation in the spherical quantum state 
with good parity one would have
to resort to two-body information or correlation functions,
cf. Fig.~(2) in Ref.~\cite{neff:2003}.

\subsection{Phenomenological interaction}
\label{Sec:sugimoto}

In Refs.~\cite{STI03,IST04,SIT04a,sugimoto,SIT07} 
Ikeda, Sugimoto and Toki  propose the idea to
mix proton and neutron wave function at the single-particle level and then perform  charge 
and parity projection of the many-body state. They use for the 4 single-particle
states for $^4$He ($\nu=1,2;\ m=\pm1/2$)
\begin{equation}
\braket{\vek{x}}{\nu, \half m} \!=\!\!
\sum_{{m_t=p,n}\atop{l=0,1}} \!
 \phi_{\nu,lm_t}\! (x) \left[Y^l (\theta, \phi)\ket{\half}\right]^{\frac{1}{2}}_m 
\otimes\ket{m_t} \ .
\label{sugimoto_wf}
\end{equation}
Here, the spatial part of the wave function contains  $s$- and $p$-wave 
components by construction, and the radial part $\phi_{\nu,lm_t}$ is expanded in terms of 
Gaussian functions.
A phenomenological interaction that is composed of a central Volkov potential 
\cite{Volkov} and the tensor plus spin-orbit  G3RS force \cite{G3RS} is used. 
It is argued that the mixing of parity and isospin is able to account for appropriate tensor 
correlations in the $\alpha$-particle when the triplet-even component of the  
central part is reduced by 0.81 and the $\op{\tau}_i \op{\tau}_j$
part of the tensor force enhanced by a factor 1.5. For this special interaction,
which we will refer to as Sugimoto-Ikeda-Toki (SIT) interaction, the expectation value of 
the tensor potential amounts to $-30$ MeV and the L=2 admixture to 7.3 \%
in their 4-body state.

\begin{table}
\caption{Comparison of expectation values of the Hamiltonian, the intrinsic
kinetic energy, the total potential energy (Coulomb included) 
and the tensor potential when using the trial states of Ref.~\cite{sugimoto},
charge and parity projected FMD and an exact calculation. For
all cases the SIT-potential \cite{sugimoto} is employed. 
Numerical values are in MeV.}
\begin{ruledtabular}
\begin{tabular}{crrrc}
{}&{Ref.~\cite{sugimoto}}&{FMD-CPP} &{Exact}\\
\hline
$\expect{\op{H}}$   &  -28.19 & -35.44  & -121.77 & \\
$\expect{\op{T}}$   &   64.39 &  63.78  &  150.56 & \\
$\expect{\op{V}}$   &  -92.58 & -99.22  & -272.33 & \\
$\expect{\op{V}_T}$ &  -30.59 & -35.19  & -207.08 & \\
\end{tabular}
\end{ruledtabular}
\label{table4}
\end{table}

Using the FMD basis with two Gaussians per nucleon, in order to allow for a $p$-wave 
component in the single-particle wave function, we obtain the results presented in 
Table \ref{table4}, which correspond approximately to those in Ref.~\cite{sugimoto}.
But the tensor contribution in the FMD state
is 4.5~MeV larger and the ground state is by 7.25 MeV more bound, which means that
the FMD state is a better variational state and represents more tensor correlations.
In contrast to that, as seen in section \ref{sec:correlationenergy},
the phase-shift equivalent and in this sense realistic potential $V_{\rm UCOM}$ 
leads typically to only -5~MeV tensor contribution with the same type of FMD trial state. 

This suggests that the SIT-potential has an unrealistic ratio of tensor to 
central potential. To investigate that further we perform an exact calculation
using an HH expansion and a Lee-Suzuki transformation to accelerate convergence, as  
proposed in \cite{BaL00}.
The exact results shown in Table~\ref{table4} exhibit a dramatic overbinding with
the SIT-potential. The reason is that the tensor potential gives a completely 
unrealistic contribution of -207~MeV. Thus neither the FMD nor the trial state of
Ref.~\cite{sugimoto} can represent in a reliable way the huge tensor correlations
induced by the unrealistic phenomenological tensor part of the SIT-interaction.

\begin{figure}[t]
\begin{center}
\includegraphics*[scale=0.6]{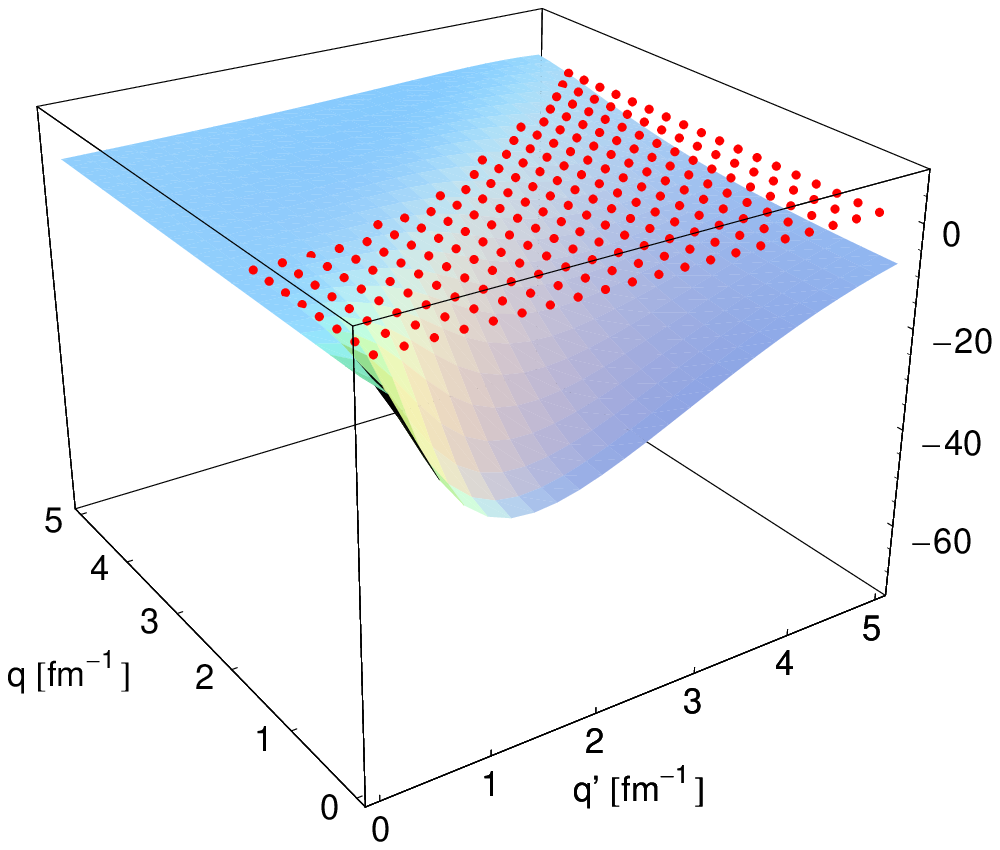}
\includegraphics*[scale=0.6]{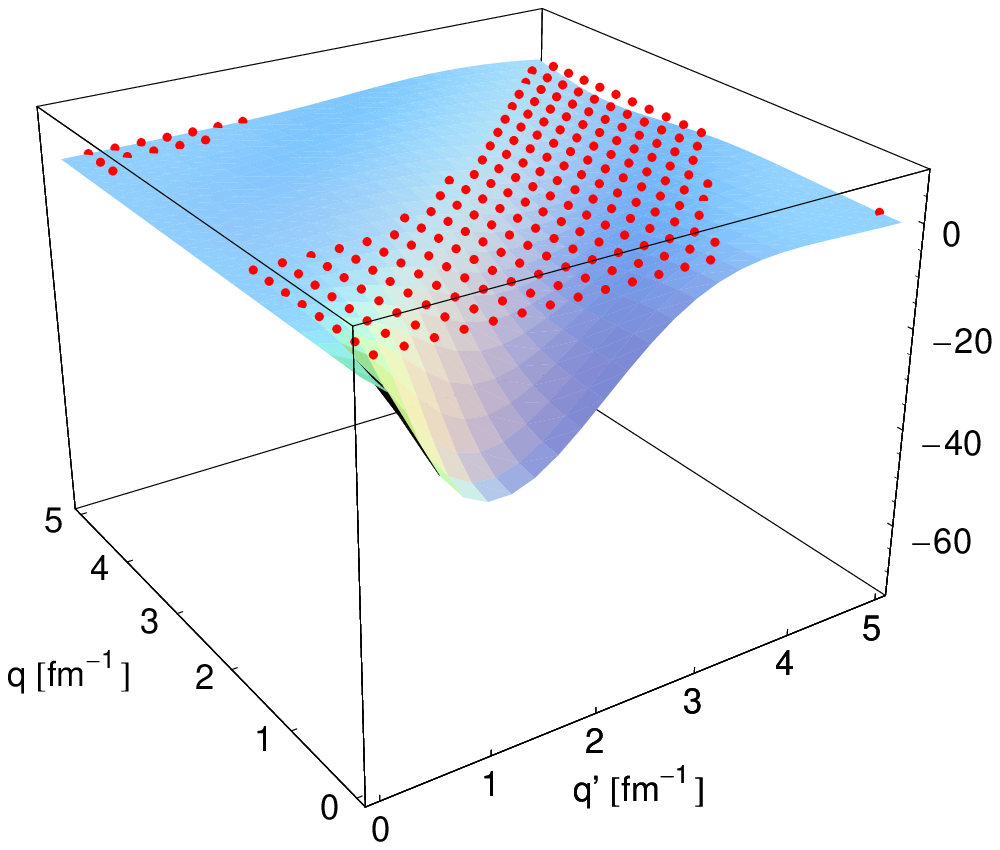}
\includegraphics*[scale=0.6]{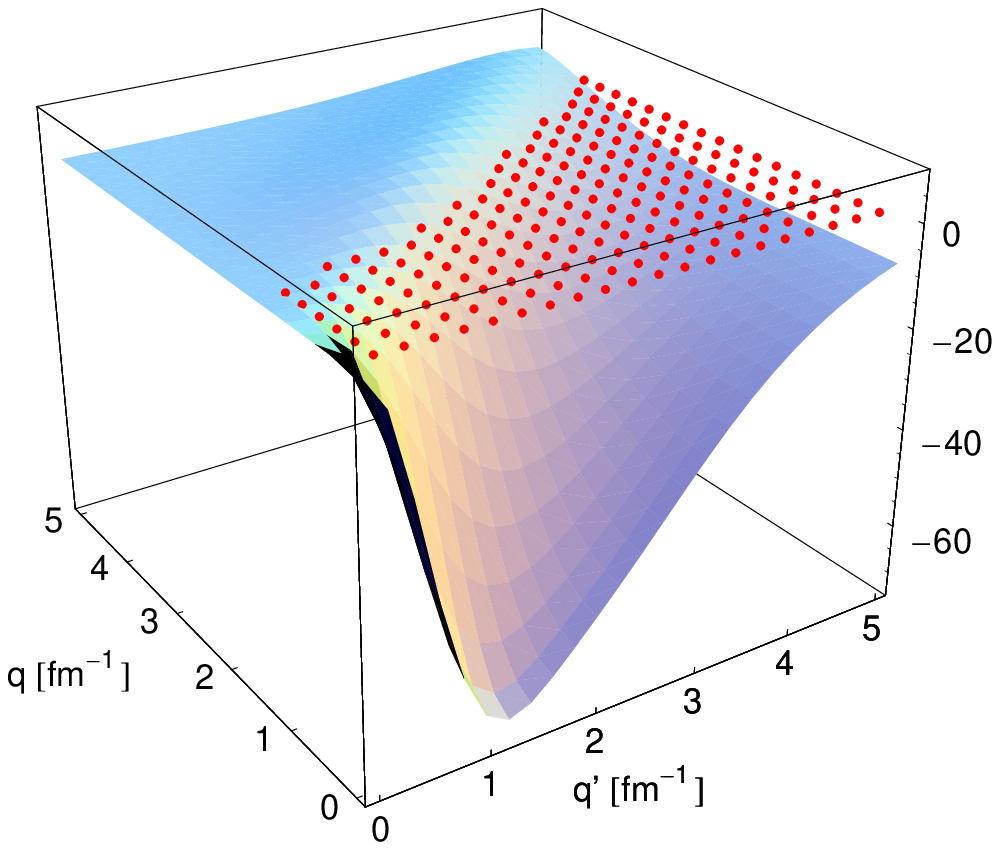}
\caption{(Color online) 
Off-diagonal momentum space matrix elements (in MeV~fm$^3$) of  
the initial AV18 potential (top),
the $V_{\rm UCOM}$ potential for $I_\vartheta=0.09$~fm$^3$ (center),
and the SIT-potential (bottom)
between the $^3S_1$ and $^3D_1$ channel.
Dotted plane indicates 0~MeV~fm$^3$
\label{figure8}}
\end{center}
\end{figure}

The reason for the large difference between the FMD or the Sugimoto {\it et al.} 
results \cite{sugimoto} 
and the exact result becomes understandable when looking at Fig.~\ref{figure8},
where the non-diagonal momentum space matrix elements of the 
AV18 potential, the corresponding $V_{\rm UCOM}$
for $I_\vartheta=0.09$~fm$^3$, and 
the SIT potential between the $^3S_1$ and $^3D_1$ channel, 
that is sensitive to the tensor force only, are displayed.
The off-diagonal matrix elements of the initial interaction are strongly reduced 
for \mbox{$q' \gtrsim 2\rm{~fm}^{-1}$} by the UCOM transformation \cite{roth:2005}.
In contrast to that the SIT matrix elements are not only much larger than 
those of $V_{\rm UCOM}$ but even larger than those of the  
AV18 interaction and the SIT interaction connects low momenta with
momenta high above the Fermi momentum.
Those high momentum components above about 2~fm$^{-1}$ are not present in a low-momentum 
basis consisting of a few Slater determinants and thus the according
correlations or admixtures cannot be represented.
On the other side the exact HH state includes high momenta and thus can
accommodate the correlations caused by this part of the potential.
This explains the drastic differences seen in the energies.
 
The SIT-interaction is an example that fitting a phenomenological
potential to a specific nucleus without reference to realistic interactions
can lead to unstable and unpredictable results and does not allow to draw definitive 
conclusions on tensor correlations in nuclei. 
States that live in a too small Hilbert space 
are forced by too strong interactions to produce the desired kind of correlations. 
Ikeda, Sugimoto and Toki are aware of this problem and argue in
some of their papers with the strength of the G-matrix \cite{IST04, sugimoto, SIT07}.
But as $V_{\rm UCOM}$ plays the same role as a G-matrix we believe that the 
long range tensor correlations can only partly be represented by
a charge and parity projected intrinsic Slater determinant, as shown
in Sec.~\ref{sec:correlationenergy}.

\subsection{Time reversal symmetry}
In this section we would like to make a remark about time reversal symmetry
\cite{Bohr-Mottleson} of our charge and parity projected FMD state.
With the charge and parity projection we actually create intrinsic states which 
are not invariant under time reversal.
Since the variational single-particle parameters are
complex they may get a non-zero imaginary part, violating the time reversal symmetry.
In order to restore the symmetry, we generate a time  reversal symmetric state by
\begin{equation}
\ket{\Psi} =
\op{P}_{CM}\op{P}^J_{MK} \op{P}^{\pi} \op{P}^{M_T}\left(
 e^{i\phi} \ket{Q} + e^{-i\phi} \ket{\bar{Q}}\right) \ ,
\end{equation}
where  $\ket{\bar{Q}} = \op{\mathcal{T}} \ket{Q}$, is the time 
reversed Slater determinant. Then we minimize the energy with respect to the phase $\phi$;
\begin{equation} 
E_0=\min_{\{\phi \}} ~
       \frac{\matrixe{\Psi}{\op{H}}{\Psi}}{\braket{\Psi}{\Psi}}  \ .
\end{equation}
We find that for the investigated cases the effect of the symmetry restoration is at most 
of the order of 100~keV and thus negligible for the purpose of this paper.

\section{Conclusions}
\label{Sec:Conclusions}

Within Fermionic Molecular Dynamics the effects of mixing proton and neutron 
components of the single-particle states of a single Slater determinant are investigated.
For that we perform for $^4$He variational calculations by minimizing the energy of 
the charge and parity projected Slater determinant
using realistic nucleon-nucleon interactions.
It turns out that the variation needs to be performed after charge and parity projection
in order to obtain a non-vanishing tensor contribution to
the ground state energy of the doubly magic nucleus $^4$He.
The intrinsic state that is not yet charge projected 
features a non-vanishing pseudo-vector iso-scalar 
spin-isospin density that is intimately related to the pseudo-scalar
iso-vector pion fields which are responsible for the long range
part of the tensor force.

A Hartree-Fock type variation without projection does not break
charge and parity of the single-particle states and hence the 
expectation value of the tensor interaction is zero.

The extra correlation energy obtained by the new degrees of freedom
that mix charge turns out to be small for realistic interactions,
smaller than anticipated from earlier work by Sugimoto, Ikeda and Toki.
The main reason is that they did not use a realistic interaction
but adopted a tensor force that scatters to high momenta and at low
momenta is about twice the strength of the tensor part in  
$V_{\rm UCOM}$.

Our result is that charge mixing and parity breaking of one-body states can account
only for a fraction of the long range correlation energy missing in 
a mean-field picture (single Slater determinant).
One should however keep in mind that $^4$He is the most demanding
nucleus in this respect. Even if the additional energy 
due to long range correlations is found to be small here,
we cannot extrapolate this result to open shell nuclei.
Also a variation after angular momentum projection might
change the situation. These issues will be subject of future 
investigations.

\begin{acknowledgments}
One of the  authors (S.B.) would like to thank S.~Quaglioni for useful 
discussion about the no-core shell model. We are grateful to P.~Navr\'atil
and N.~Barnea for providing us with the NCSM and HH codes, respectively. 
\end{acknowledgments}


\end{document}